\documentclass[letterpaper, 10 pt, conference]{ieeeconf}  

\IEEEoverridecommandlockouts                              

\usepackage{graphics} 
\usepackage{amsmath} 
\usepackage{amssymb}  
\usepackage{optidef}
\usepackage{mathrsfs}
\usepackage{siunitx}
\usepackage[dvipsnames]{xcolor}
\usepackage{mathptmx}

\usepackage{xcolor}
\usepackage[dvipsnames]{xcolor}
\usepackage{booktabs}
\usepackage{multirow}
\usepackage{subfig}
\usepackage{enumerate}
\usepackage{siunitx}  
\usepackage{physics}

\usepackage{graphicx}      

\usepackage[english]{babel}
\newcommand{\ps}{\textcolor{black}}
\newcommand{\ks}{\textcolor{black}}
\newcommand{\st}[1]{\mathbf{#1}}
\newcommand{\nn}[1]{\mathsf{#1}}

\DeclareMathOperator*{\argmin}{arg\,min}

\title{\LARGE \bf
Deep Neural Network Based Optimal Control of Greenhouses
}

\author{Kiran Kumar Sathyanarayanan$^{1}$, Philipp Sauerteig$^{1}$, and Stefan Streif$^{1,2}$
\thanks{The authors are with $^{1}$Technische Universit\"at Chemnitz, 09126 Chemnitz, Germany, Automatic Control and System Dynamics Lab and $^{2}$Fraunhofer Institute for Molecular Biology and Applied Ecology, Department of Bioresources, Giessen, Germany  (e-mail: \{kiran.sathyanarayanan,philipp.sauerteig,stefan.streif\}@etit.tu-chemnitz.de)}
}

\begin{document}

\maketitle
\thispagestyle{empty}
\pagestyle{empty}

\begin{abstract}

Automatic control of greenhouse crop production is of great interest owing to the increasing energy and labor costs.
\ks{In this work, we} use two-level control, where the upper level generates suitable reference trajectories \ks{for states and control inputs} based on day-ahead predictions.
These references are tracked in the lower level using Nonlinear Model Predictive Control (NMPC).
In order to apply NMPC, a model of the greenhouse dynamics is essential. 
However, the complex nature of the underlying model including discontinuities and nonlinearities results in intractable computational complexity and long sampling times.
As a remedy, we employ NMPC as a data generator to learn the tracking control policy using deep neural networks. 
Then, the references are tracked using the trained Deep Neural Network (DNN) to reduce the computational burden.  
The efficiency of our approach under real-time disturbances is demonstrated by means of a simulation study. 

\end{abstract}


\section{Introduction}\label{sec:intro}
The world population is projected to increase from 7 billion in 2010 to 9.8 billion in 2050, resulting in a 50$\%$ increase in food demand~\cite{searchinger2019creating}. Despite the need to increase food production, it cannot be achieved by expanding agricultural land due to the challenge of ever-growing urban space and climate change.
Controlled Environment Agriculture (CEA) like greenhouse farms and indoor vertical farms have gained attention due to their capability to produce high yield in limited space.
However, high energy spikes and the need for energy-efficient operations have posed a question on the environmental impact and sustainable operation of CEA~\cite{ijerph20021116}.  

The usage of renewable energy and the development of solar greenhouses is a viable option to achieve green and sustainable food production~\cite{HASSANIEN2016989}.
With frequently shifting rules and unpredictable energy prices, automatic control systems are the most efficient alternative for greenhouse climate control and crop growth.
Several approaches to improve energy efficiency in greenhouses, reduce CO$_2$ emissions, and enhance water use efficiency have been proposed~\cite{IDDIO2020109480, REN2022157756}. 
Classical control approaches to maintain the pre-defined greenhouse climate are discussed in~\cite{hu2011,Su2016772}.

Optimal control approaches using mathematical models to maintain greenhouse climate (temperature and humidity) for plant production demonstrating their economical benefits are shown in~\cite{VANBEVERENmin}.
Here, Hierarchical Model Predictive Control (HMPC) is employed where the states are divided into different timescales, and the control system is designed for each levels to maintain the greenhouse climate. 
The upper-level control system generates the optimal climate reference trajectories using the greenhouse model and the weather forecast data, whereas the lower-level control system controls the fast-changing climate by tracking the reference. 
In~\cite{RAMIREZARIAS2012490}, multiobjective optimization is proposed to achieve profitable crop growth along with maintaining crop quality and irrigation efficiency using the model of greenhouse and crop growth. 
Optimal control of a solar greenhouse with an economic objective to increase the yield and decrease the gas usage is discussed in~\cite{van2007optimal}. 
In general, a longer prediction horizon is considered to generate an optimal control input \ks{in NMPC. 
	Assuming a larger sampling time reduces computational demand for longer predictions.}

\ks{The system equations governed by NMPC are continuous over time and need to be discretized for implementation on embedded or similar devices. 
	The choice of sampling time significantly influences accuracy, making shorter sampling times preferable.
	\ps{However, the discretization method employed to implement optimal control using greenhouse climate and crop models, as well as the impact of sampling time on integration accuracy are not commonly discussed.}}
The effect of small sampling time to be chosen to adequately approximate the system behavior on a given prediction horizon is dealt with in~\cite{PADMANABHA202015771}.
However, the smaller sampling time demands a shorter horizon for sufficiently fast computation of nonlinear optimization problems. 
In~\cite{LUCIA2018511, PARISINI19951443, PONKUMAR2018512}, the problem of higher computation demand of NMPC due to longer prediction horizons has been overcome by the use of DNNs to generate an approximate solution to the NMPC problem. 
\ks{The universal approximation property of the DNNs has motivated the usage of machine learning for optimal control \cite{PARISINI19951443,learningHERTNECK}.}

In this paper, we propose to design a two-level hierarchical control for a semi-closed solar collector greenhouse. 
The upper-level optimization layer generates the economically optimal input and state reference trajectory for a single day using the weather forecast data, while the lower-level NMPC tracks the above reference trajectory on satisfying the desired bounds with a short-term disturbance forecast.
The main advantage is that the approximated \ks{NMPC} can be deployed in low-cost embedded hardware, which is more desirable in remote locations like refugee camps.
Indeed, the approximate DNN based controller can be applied to any CEA system, where fast computation is desirable.

First, a mathematical model describing the greenhouse climate dynamics and crop growth is presented \ks{in Section~\ref{sec:model}.} 
To describe the plant growth, we use the tomato model presented in~\cite{van2007optimal}.
\ks{Following that, Section~\ref{sec:ocp} introduces the two-level hierarchical control of the greenhouse.
	Then, Section~\ref{sec:DNN} discusses the training of DNN to approximate the optimal control inputs given the current state, reference and disturbances.}
\ks{Finally, a two-level control of a greenhouse with a lower level of control using trained DNN is implemented and its performance is reviewed in Section~\ref{sec:sim}, with the conclusion in Section~\ref{sec:conclusion}.} 


\section{Greenhouse and Crop Model}\label{sec:model}

This section summarizes the mathematical model of the greenhouse climate along with the crop. 
For our study, a Venlo-type greenhouse growing tomato crop fitted with a blackout screen, vent opening through the roof, CO$_2$ injection, heating, and cooling source is considered. 
In this approach, the greenhouse climate incorporates temperature ($T$), carbon dioxide concentration ($C$), and absolute humidity ($H_\mathrm{a}$) of the air inside the greenhouse. 
A mathematical model of the greenhouse climate is derived via the mass and energy balance equations, see also~\cite{VANBEVERENmin,VANBEVERENopt,bontsema2008line}. 

\subsection{Dynamic Model of the Greenhouse Climate}\label{subsec:dyn_model_gh}
\subsubsection{Temperature} 
The temperature $T$ of greenhouse air is influenced by various heat fluxes and the energy balance equation describing the same is given by 
\begin{equation}\label{eq:temp}
	\dot T = \left(Q_\text{sun} + Q_\text{vent} + Q_\text{cov} + Q_\text{trans} + Q_\text{heat} - Q_\text{cool}\right) /k_\text{C,gh} ,
\end{equation}
where $k_\text{C,gh}$ is the total heat capacity of the greenhouse per unit area (\si{\joule \per \meter \squared \per \kelvin}). 
Moreover, \ks{$Q_\text{sun}  = k_\text{tot}\cdot Q_\text{rad} \,$ denotes the heat flux term contributed by the incoming radiation in \si{\watt \per \meter \squared},
	where $k_{tot}$ is the total transmittance of the cover.} 
The heat flux component due to vent opening is given by
\begin{equation*}\label{eq:qvent}
	Q_\text{vent} = k_\text{c,air} \cdot k_{\rho,\text{air}}\cdot k_\text{u,vent} \cdot u_\text{V} (T_\text{out} - T) \quad  [\si{\watt \per \meter \squared}],  
\end{equation*}
where $k_\text{c,air}$ and $k_{\rho,\text{air}}$ denoting the specific heat capacity and density of the air, respectively, and $k_\text{u,vent}$ is the maximum ventilation of the greenhouse.
Leakage through the cover is assumed to be negligible. 
The convective heat loss through the cover of the greenhouse is given by 
\begin{equation*}\label{eq:qcov}
	Q_\text{cov} = k_\text{A} \cdot k_\text{U} \cdot (T_\text{out} - T) \quad [\si{\watt \per \meter \squared}], 
\end{equation*}
where $k_\text{U}$ is the coefficient of heat transfer through the cover, $k_\text{A}$ is the ratio of the surface area of the cover and the greenhouse floor. 

We consider the transpiration of tomato crops in a greenhouse given by~\cite{bontsema2008line}. 
Here, $Q_\text{trans}$ is the heat energy absorbed due to crop evapotranspiration,and can be obtained by
\begin{equation*}\label{eq:qtrans}
	Q_\text{trans} = g_\text{e} k_\text{L} (H_\text{v} - x_\text{P,H})  \quad [\si{\watt \per \meter \squared}], 
\end{equation*}
where $g_\text{e}$ is the transpiration conductance, $k_\text{L}$ is the amount of energy needed to evaporate water from a leaf, and $H_\text{v}$ is the absolute water vapor concentration at crop/vegetation level. 
\ks{The quantities are computed as follows. The transpiration conductance is $g_\text{e} = 2 k_\text{LAI} / \left((1 + \epsilon) r_\text{b} + r_\text{s} \right) \, [\si{\meter \per \second}],$}
where $k_\text{LAI}$ is the leaf area index, $\epsilon$ is the ratio of latent to sensible heat content of saturated air for a change of $1 \si{\degreeCelsius}$ in $T$, $r_\text{b}$ is the resistance to heat transfer of the leaf boundary layer, and $r_\text{s}$ is the stomatal resistance. 
Moreover, $H_\text{v}$ is calculated as
\begin{equation*}\label{eq:Hv}
	H_\text{v} = H_\text{sat} + \epsilon r_\text{b} / (2k_\text{LAI}) \cdot R_\text{n} / k_\text{L} \quad [\si{\gram \per \meter \cubed}],
\end{equation*}
where $H_\text{sat}$ is the saturated vapor concentration of the greenhouse air that can be approximated by, $	H_\text{sat} = 5.5638 \, \mathrm{e}^{0.0572 T}$
for $T \in [10,40].$
The net radiation $R_\text{n}$ at crop level, and its relation with the global radiation $Q_\text{rad}$ is given by
\begin{equation*}\label{eq:Rn}
	R_\text{n} = 0.86(1 - \mathrm{e}^{-0.7 k_\text{LAI}}) Q_\text{rad} \quad [\si { \watt \per \meter \squared}].
\end{equation*}
The heating and cooling power $Q_\text{heat}$ and $Q_\text{cool}$, respectively, in $\si{\watt \per \meter \squared}$ are given by
$Q_\text{heat} = {u_{\text{Q}_\text{h}} \cdot\text{P}_\text{h,max}}$ and $Q_\text{cool} = {u_{\text{Q}_\text{c}} \cdot \text{P}_\text{c,max}}$,
where $\text{P}_\text{h,max}$ and $\text{P}_\text{c,max}$ are the maximum heating and cooling power per unit greenhouse floor area, respectively.

\subsubsection{Carbon-dioxide concentration}
The mass balance equation of the CO$_2$ concentration is described as 
\begin{equation}\label{eq:co2}
	\dot C = k_\text{A,gh} / k_\text{V,gh} \cdot \left( C_\text{inj} + C_\text{vent} - C_\text{phot} \right),
\end{equation} 
where $k_\text{V,gh}$ is the volume of the greenhouse.
The pure industrial CO$_2$ injection rate $C_\text{inj} = u_{\text{C}} \cdot C_\mathrm{max} / k_\text{A,gh}  \, [\text{g}\, \text{m}^{-2} \text{s}^{-1}].$ 
The CO$_2$ exchange due to ventilation $C_\text{vent}$ is described as
\begin{equation*}\label{eq:cvent}
	C_\text{vent} =  k_\text{u,vent} \cdot u_\text{V} (C_\text{out} - C) \quad [\text{g}\, \text{m}^{-2} \text{s}^{-1}].
\end{equation*}
Finally, the net CO$_2$ consumption due to photosynthesis and transpiration by the crop is given by 
\begin{equation*}
	C_\text{phot} = \phi_{\text{CO}_2} \cdot 10^{-3} / k_\text{A,gh} \quad [\text{g}\, \text{m}^{-2} \text{s}^{-1}],
\end{equation*}
where $\phi_{\text{CO}_2}$ is defined in Section~\ref{subsec:crop}.

\subsubsection{Absolute humidity}
The absolute humidity of the greenhouse air is influenced by various vapor fluxes and the mass balance equation reads as
\begin{equation}\label{eq:hum}
	\dot H_\mathrm{a} =  k_\text{A,gh}( H_\text{trans} - H_\text{vent} - H_\text{cov} + H_\text{heat} - H_\text{cool} ) / k_\text{V,gh}. 
\end{equation}
The amount of water vapor produced due to plant transpiration $H_\text{trans} = g_\text{e} (H_\text{v} - H_\mathrm{a}) \, [\text{g}\, \text{m}^{-2} \text{s}^{-1}].$
Condensation of vapor to the cover $H_\text{cov}$ occurs when the temperature of the cover $T_\text{cvr}$ is below the dew point temperature of the air, and it is calculated as 
\begin{equation*}\label{eq:hcov}
	H_\text{cov} = g_\text{C} \left(0.2522 \mathrm{e}^{0.0485 T}(T \!-\! T_\text{out}) \!-\! (H_\text{sat} \!-\! H_\mathrm{a})\right) \, [\text{g}\, \text{m}^{-2} \text{s}^{-1}].
\end{equation*}
Here, condensation conductance $g_\text{C}$ is described in~\cite{STANGHELLINI1995129} as,
\begin{equation*}\label{eq:gc}
	g_\text{C} = \text{max}(0, {k_\text{A}} \, 1.64\times 10^{-3} (T - T_\text{cvr})^{\frac{1}{3}}) \quad [\si{\meter \per \second}].
\end{equation*}
The vapor flux due to ventilation $H_\text{vent}$ and direct air heater $H_\text{heat}$ are described as
\begin{equation*}\label{eq:hvent}
	H_\text{vent} = k_{u,vent}\cdot u_\text{V}  (H_\mathrm{a} - H_\text{out}), \quad 
	H_\text{heat} = k_\eta Q_\text{heat} \quad [\text{g}\, \text{m}^{-2} \text{s}^{-1}],
\end{equation*}
where $k_\eta$ $[\si{\gram \per \joule}]$ is the amount of vapor released when 1~\si{\joule} of sensible energy is produced by the heater~\cite{vanthoor2011model}. 
The cooling of air happens through the heat exchanger, where the pipe inlet is maintained at a constant temperature. 
The condensation of vapor on the pipes $H_\text{cool}$ is given by
\begin{equation*}
	H_\text{cool} = \text{max}(0,u_{\text{Q}_\text{h}} k_\text{A,pipe} \cdot k_\text{cool} (H_\mathrm{a} - H_\text{sat,cool})),
\end{equation*}
as described in~\cite{van2007optimal}. 
Here, $k_\text{A,pipe}$ is the surface area of the cooling pipe, $k_\text{cool}$ is the mass transfer coefficient of water vapor from indoor air to the cooling tube, and $H_\text{sat,cool}$ is the saturation concentration of water vapor at the temperature of the cooling pipe.

\subsection{Crop Model}\label{subsec:crop}
Photosynthesis is the process through which crops assimilate carbon dioxide and water to support their maintenance, growth, and development. This process is affected by higher radiation intensity and elevated CO$_2$ concentration~\cite{dannehl10122808}. 
Here, the CG4 photosynthesis model presented in~\cite{van2007optimal} is used. 
Furthermore, it is assumed that required water and nutrients for crop growth are available at all times. 
This model describes the net CO$_2$ consumption rate of the crop as a function of outdoor radiation $Q_\text{rad}$, CO$_2$ concentration~$C$, temperature of the crop, and humidity~$H_\mathrm{a}$. 
As the temperature of the crop is not measured, temperature~$T$ of the greenhouse air is used.
The consumed CO$_2$ are converted into carbohydrate assimilate, resulting in the biomass production
\begin{equation}\label{eq:bio}
	\dot B =  (k_{\text{B,CO}_2} \phi_{\text{CO}_2}) / k_\text{A,s} \quad [\text{kg}\, \text{m}^{-2} \text{s}^{-1}].
\end{equation}
Here, $B$ is the fresh biomass weight, $\phi_{\text{CO}_2}$ is the net photosynthesis rate of the canopy, and $k_{\text{B,CO}_2}$ is the conversion factor to convert the consumed CO$_2$ to the fresh weight biomass. 


\section{Optimal Greenhouse Control}\label{sec:ocp}

The dynamics of the greenhouse climate and the crop growth were summarized in ~\eqref{eq:temp},~\eqref{eq:co2},~\eqref{eq:hum}, and~\eqref{eq:bio}.
Four control inputs are defined to maintain the states - temperature $T$, CO$_2$ concentration $C$, and relative humidity $H$ of the greenhouse air within the desired bounds while minimizing the production costs for high yields using a cost function.

The system behavior is of the form
\ks{\begin{equation}\label{eq:ss}
		\st{\dot{x}}(t) = {f(\st{x}(t),\st{u}(t),\st{d}(t))}
	\end{equation}
	with state vector $\st{x} = \begin{bmatrix}
		T & C&  RH&  B
	\end{bmatrix}^\top$, input vector $\st{u} = \begin{bmatrix}
		u_\text{V} & u_\text{C} & u_{\text{Q}_\text{h}} & u_{\text{Q}_\text{c}}
	\end{bmatrix}^\top,$ and disturbance vector $\st{d} = \begin{bmatrix}
		T_\text{out} & C_\text{out} & H_\text{out} & Q_\text{rad}
	\end{bmatrix}^\top.$}

The model represented by~\eqref{eq:ss} exhibits discontinuities in the form of maximum operators. 
To enable various options for optimization algorithms, the maximum operators are approximated as described in~\cite{biswasSmooth}, and is given as $\text{max}(a,b) \approx \left(a + b + \sqrt{(a-b)^2 + \gamma}\right)/2,$
where $\gamma = 10^{-4}$ is a tuning parameter. 

In the following, a framework for the optimal control of the greenhouse is implemented.
As the task of observer design to get the biomass from other measurements in outlook is omitted due to page brevity, full state availability at every time instance is assumed.
The framework includes two levels of control: an open-loop optimization layer and a closed-loop reference tracking Nonlinear Model Predictive Control (NMPC) layer.
The upper level solves an economic objective function with the forecast weather data to generate a reference trajectory for the NMPC. 
Subsequently, the NMPC tracks the obtained reference trajectory in real time.

Optimization problem and predictive control is performed at discrete time step, for which the dynamics 
\ks{\begin{equation}
		\st{x}(t_{k + 1}) = \st{x}(t_k) + \Delta t \cdot f(\st{x}(t_k),\st{u}(t_k),\st{d}(t_k))
	\end{equation}
	are utilized, where $t_k = n_\mathrm{o}\Delta t_\mathrm{o}$, $n_\mathrm{o} \in \mathbb{N}_0$, and $\Delta t_\mathrm{o} > 0$ is the sample time for the open loop with $\nn{T}$ being the total simulation time.}
Similarly, the closed-loop system is discretized, where $t_k = n\Delta t_\mathrm{c}$, $n \in \mathbb{N}_0$, and $\Delta t_\mathrm{c} > 0$ is the sample time, with $\Delta t_\mathrm{o} > \Delta t_\mathrm{c}$.
The discretization is performed using orthogonal collocation of degree~4. 

\subsection{Open-loop control}
The objective is to 
maximize crop production while minimizing the total operating costs which include costs for heating/cooling, ventilation, and CO$_2$ injection.
To this end, we define the day-ahead cost function 
\ks{\begin{equation}\label{eq:objOCP}
		J_\mathrm{o}{\color{black}(\st{u})} = {\sum_{k=0}^{N_\mathrm{o} - 1} L(\st{x}(t_k),\st{u}(t_k)) - V(\st{x}(t_{N_\mathrm{o}}))}
	\end{equation}
	with ${N_\mathrm{o}} = \frac{\nn{T}}{\Delta t_\mathrm{o}} \in \mathbb{N}^+$. 
	The stage costs $L(\st{x}(t_k),\st{u}(t_k))$ and terminal costs $V(\st{x}(t_{N_\mathrm{o}}))$ 
	and are given by  
	\begin{IEEEeqnarray}{rCl}
		L(\st{x}(t_k),\st{u}(t_k)) & = & p_\text{E} \cdot P_\text{v}^\text{max} \cdot u_{\text{V}}(t_k) 
		+ p_\text{C} \cdot C_{\text{inj}}^{\text{max}} \cdot u_{\text{C}} (t_k) \nonumber\\
		&& + p_\text{E} \cdot P_{\text{Q}_\text{h}}^{\text{max}} \cdot u_{\text{Q}_\text{h}}(t_k)
		+ p_\text{E} \cdot P_{\text{Q}_\text{c}}^{\text{max}} \cdot u_{\text{Q}_\text{c}}(t_k) \nonumber\\
		&& + \nn{P}_\mathnormal{T}(\st{x})  + \nn{P}_\mathnormal{C}(\st{x})
		+ \nn{P}_\mathnormal{RH}(\st{x})\nonumber\\
		V(\st{x}(t_{N_\mathrm{o}})) & = & p_\text{y} \cdot k_{\text{B}_\text{fruit}} \cdot k_\text{A,s} \cdot B(t_k). 
		\label{eq:rcost}
	\end{IEEEeqnarray}
	Here, penalties for temperature $\nn{P}_\mathnormal{T}(\st{x})$, relative humidity $\nn{P}_\mathnormal{RH}(\st{x})$ and CO$_2$ concentration $\nn{P}_\mathnormal{C}(\st{x})$ are defined as
	\begin{equation}\label{eq:pen}
		\nn{P}_\st{x}(\st{x}) = \begin{cases}
			c_\st{x} \cdot |\st{x}^\text{min} - \st{x}|,& \text{if } \st{x} < \st{x}^{\text{min}}\\
			0,              & \text{if } \st{x}^{\text{min}} \leq \st{x} \leq \st{x}^{\text{max}}\\
			c_\st{x} \cdot |\st{x}^\text{max} - \st{x}|,& \text{if } \st{x} > \st{x}^{\text{max}},
		\end{cases}
	\end{equation}
	where $c_\st{x}$ is the weight factor associated with the state $\st{x}$ exceeding the bounds $\st{x}^{\text{min}}$ and $\st{x}^{\text{max}}$. 
	With the penalty function increasing linearly in value with respect to the deviation, the discontinuities in~\eqref{eq:pen} are smoothed via
	\begin{equation}\label{eq:spen}
		\begin{split}
			\nn{P}_\st{x}(\st{x}) & \approx \frac{c_\st{x}}{2} \cdot \left( \sqrt{(\st{x}^\text{min} - \st{x})^2 + \gamma} +  \sqrt{(\st{x}^\text{max} - \st{x})^2 + \gamma} \right. \\ 
			& \qquad \left. -  ( \st{x}^\text{max} - \st{x}^\text{min}) \right) .
		\end{split} 
\end{equation}}
These penalty functions are used as soft constraints. 
Economic profit is represented by a negative cost term, which is calculated from the difference between running and terminal cost.
In~\eqref{eq:rcost}, $p_\text{E}$ is the price of electricity, $p_\text{C}$ is the price of industrial CO$_2$, $p_\text{y}$ is the selling price of yield, and $k_{\text{B}_\text{fruit}}$ is the percentage of biomass corresponding to fruit.
The operating cost of the vent is obtained by multiplying the maximum power rating of the ventilation motor $P_\text{v}^\text{max}$, $p_\text{E}$ and the input $u_{\text{V}}$.
Likewise, the cost of CO$_2$ injection and the operating cost of heating and cooling are obtained. 
We restrict inputs $\st{u}(t) \in \mathcal{U}$ and states $\st{x}(t) \in \mathcal{X}$ in order to maintain temperature, CO$_2$ concentration, and humidity within the safe limit, despite being penalized as soft constraints. 
These considerations motivate the Optimal Control Problem (OCP)

\ks{\begin{align}
		\st{u}^*_\mathrm{o} = \argmin_\st{u} \quad & J_\mathrm{o}(\st{u})\label{eq:OLP}\\
		\mathrm{s.t.} \quad & {\st{x}t_{k+1})} = {\st{x}(t_k) + \Delta t_\mathrm{o} \cdot \st{f}(\st{x}(t_k),\st{u}(t_k),\st{d}(t_k)), \nonumber} \\
		& \st{x}(t_0) = \st{x}_0, \nonumber \\
		& {\st{u}(t_k)} \in  \mathscr{U} \forall k \in \{0,.....,N_\mathrm{o}-1\}, \nonumber\\
		& {\st{x}(t_k)} \in  \mathscr{X} \forall k \in \{0,.....,N_\mathrm{o}-1\}, \nonumber
\end{align}}
where $\st{x}_0$ is the initial state and $\st{u}^*_\mathrm{o}$ is the optimal reference input for the lower-level NMPC. 

\subsection{Closed-loop NMPC}
The lower level NMPC is formulated to track the reference trajectories obtained from upper level. Nonlinear MPC is employed here to handle the sudden large deviations from reference, due to the disturbance and high nonlinearity.
Here, the optimal input variables~$\st{u}^*_\textrm{o}$ obtained from~\eqref{eq:OLP} are taken as input reference trajectories $\st{u}^\text{ref}$.
Similarly, the corresponding state variables is taken as the reference state trajectories $\st{x}^\text{ref}$ to the closed loop controller. 
The NMPC problem with a prediction horizon ${N} \in \mathbb{N}^+$ is defined by
\ks{\begin{mini!}|s|
		{{\st{u}}}{J_\mathrm{c}(\st{u}) = \, \sum_{j=i}^{i+N-1}{(\| \st{x}(t_j) - \st{x}^\text{ref}_k\|^2_Q + \|\st{u}(t_j) - \st{u}^\text{ref}_k\|^2_R)} \nonumber}{\label{eq:CLP}}{}{} 
		\breakObjective{ + \| \st{x}(t_N) - \st{x}^\text{ref}_k\|^2_P}
		\addConstraint{\st{x}(t_{j+1})} = {\st{x}(t_j) + \Delta t_\mathrm{c} \cdot \st{f}(\st{x}(t_j),\st{u}(t_j),\st{d}(t_j)) \label{eq:sysCLP}} 
		\addConstraint{\st{x}_0(t_j) = \st{x}_0}
		\addConstraint{{\st{u}(t_j)} \in  \mathscr{U} \forall j \in \{0,.....,N_s-1\}}
		\addConstraint{{\st{x}(t_j)}\in  \mathscr{X} \forall j \in \{0,.....,N_s-1\},}
	\end{mini!}
where $Q\succeq 0$, $R\succ 0$ and $P\succeq 0$ are the weighting matrices and ${N_\mathrm{s}} = \frac{\nn{T}}{\Delta t_\mathrm{c}} \in \mathbb{N}^+$. }
The NMPC problem is solved with a short-term disturbance forecast of horizon $N$.
We solve OCP~\eqref{eq:CLP} for the optimal input trajectories $\st{u}_{j}^{\star}$ and apply $\st{u}^{\star}_0$ to system~\eqref{eq:sysCLP}. 
The closed-loop NMPC feedback controller based on~\eqref{eq:CLP} can be expressed as the function
\begin{equation}\label{eq:uMPC}
	\st{u} = {\pi_\text{MPC}(\nn{w})},
\end{equation}
where $\nn{w} = \begin{bmatrix}
	\st{x} & \st{x}^\text{ref} & \st{u}^\text{ref} & \st{d}
\end{bmatrix}^\top \in \mathbb{R}^{16}$.

\section{Deep Neural Network \ks{Controller}}\label{sec:DNN}
The main idea of this paper is to approximate the reference tracking closed-loop NMPC discussed in the previous section using a DNN. 
Now, let us recall the structure of feed-forward neural networks, network training, and finally the implementation of closed-loop control using DNNs. 

\subsection{Deep Neural Networks}\label{sec:FFNN}

We aim to approximate the function~\eqref{eq:uMPC} using a feed-forward Neural Network (NN) by defining the mapping $\nn{u} = {\pi_\text{NN}\nn{(w;\Theta)}}$, where $\nn{\Theta}$ represents the unknown NN parameters.
In general, a neural network consists of one input layer, one output layer, and $N_\text{h}$ hidden layers, where an NN is called DNN for $N_\text{h}\geq2$ \cite{goodfellow2016deep}.
Moreover, it was shown in \cite{LUCIA2018511} that the performance of DNN compared to a shallow network ($N_h = 1$) results in better performance for less number of parameters $\nn{\Theta}$.

Each layer~$l$ contains $n_l$ neurons, and the hidden layers are structured as 
\begin{equation}\label{eq:nnop}
	h^l =  \beta(b^l + W^l h^{l-1}),
\end{equation}
where $h^{l-1} \in \mathbb{R}^{n_{l-1}}$ is the output of the previous layer with $h^0 = \nn{w}$. 
Here $b^l \in \mathbb{R}^{n_l}$ is called the bias vector, $W^l \in \mathbb{R}^{n_l \times n_{l-1}}$ is called the weight matrix, and $\beta$ is a nonlinear activation function. 
Typical activation functions are Rectifying Linear Unit (ReLU), tangent hyperbolic, and the sigmoid function. 
The most commonly used activation function ReLU is defined as $\beta(h) = \text{max}(0,h)$, which can enhance the training of DNN by avoiding vanishing gradients. 

The parameter $\nn{\Theta} = \{\nn{\Theta}_\mathrm{1}, \dots, \nn{\Theta}_{\mathnormal{N_h}+1} \} $ contains all the weights and biases that define the operation~\eqref{eq:nnop} of each layer. 
The number of parameters to be stored defines the memory footprint of the proposed controller and is given by $N_\nn{\Theta} = \sum_{l=1}^{N_h+1} n_l(n_{l-1} +1). $

With the architecture being defined, training of the neural networks involves finding the optimal weights~$W^l$ and biases~$b^l$ by minimizing the defined loss function.
We consider Mean Squared Error (MSE) between the optimal input ($\st{u}$) and the approximate control input ($\nn{u}$) as the loss function.

The optimization problem solved at training is given by
\begin{mini*}|s|
	{{W^l,b^l }}{\frac{1}{N_D}\sum_{m=1}^{N_D} {( \st{u}(\nn{w}^{m}) - \nn{u}(\nn{w}^{m}))^2, } }{\label{eq:MSE}}{}{}
\end{mini*}
where $N_D$ is the number of training data pairs in the training data matrix $\mathcal{D} = [\nn{w}^1,\nn{w}^2, \dots, \nn{w}^{N_D}] \in \mathbb{R}^{16 \times N_D}$, and $\mathcal{D}$ should be sufficiently large enough. 
After the network training, we use $\nn{\pi_\text{NN}(w;\Theta)}$ to infer the control input $\nn{u} \approx \nn{\pi_\text{MPC}(w)}$. 

Normalizing the training data helps to improve the numerical properties of the network when inputs to NN are in different ranges.
The normalization operation results in a normalized data matrix $\mathcal{D}_N$.
Similarly, inverse transformation is applied during the NN inference to recover the output in the range of training data.

In addition, the hyperparameters like the number $N_h$ of hidden layers, the number $n_l$ of neurons within a layer, the activation function $\beta$, and the gradient-based optimization algorithm along with its parameters like learning rate affect the performance of the network. 
Typically, experimentation is necessary to determine an optimal combination of these hyper-parameters~\cite{bergstra2012random}.

\section{Simulation and Results}\label{sec:sim}
As discussed in the previous sections, we use two-level hierarchical control to first generate economic reference trajectories for the inputs and states and, then track them using a DNN.
The OCP~\eqref{eq:OLP} is solved subject to input constraints $0 \leq \st{u} \leq 1$ and soft constraints on states with $\st{x}_\text{min} = [18, 500, 60]$ and $\st{x}_\text{max} = [26, 900, 90]$. 
The constraint on the state $C$ is generally mentioned in $\si{ppm}$. 
Additionally, 
\begin{align}
	\mathcal{X} = [14,30] \times [300, 1000] \times [10, 100] \times [0, 100] \notag 
\end{align}
defines a hard box constraint set. 
Now, the upper-level OCP is solved with sampling time $\Delta t_\mathrm{o} = 300 \, \si{\second}$ for a single day $\nn{T}=86400  \,\si{\second}$, resulting in $N_\mathrm{o} = 288$. 
On the other hand, the NMPC is solved for every $300 \, \si{\second}$ with $\Delta t_\mathrm{c} = 60 \, \si{\second}$ and prediction horizon $N=5$. 
The weighing matrices for the closed loop are chosen as $Q = 100 \cdot \mathrm{I}_4$, $R = \mathrm{I}_4$, and $P=Q$. 

\subsection{Training of the reference tracking DNN controller}
For achieving a sufficiently large data matrix $\mathcal{D}$, 8 months of disturbance data during which the greenhouse operation was active is considered.
This results in 285,000 data pairs with 80$\%$ used for training and the remainder for testing network accuracy.
\texttt{Adam}~\cite{KingmaB14a}, a stochastic gradient descent optimization method is used to perform training via \texttt{Tensorflow}~\cite{abadi2016tensorflow}.
The last timestamp values of the previous day are taken as the initial condition for the next-day optimization and reference tracking problem.

To solve the optimization problems~\eqref{eq:OLP} and~\eqref{eq:CLP}, we use \texttt{CasADi}~\cite{andersson2019casadi} with \texttt{IPOPT}~\cite{wachter2006implementation}. 
Moreover, the nonlinear dynamics are discretized using orthogonal collocation for predictive control implementation \ks{and the integration operations are performed using the \texttt{SUNDIALS} toolbox.}
In order to provide a good approximation of $\pi_\text{MPC}$ while keeping the number of parameters $N_\nn{\Theta}$ small~\cite{bergstra2012random} a random search using
\texttt{KerasTuner} is employed to determine suitable hyperparameters like $N_h$, $n_l$ and $\beta$. 
The resulting DNN includes 5 hidden layers with almost 4,000 parameters, is trained with a batch size of $512$ data pairs, a learning rate of $0.001$, and occupies less than 14~kB of memory.

\subsection{Performance of the DNN controller}
\begin{figure}[t!]
	\centering
	\includegraphics[width=8.4cm]{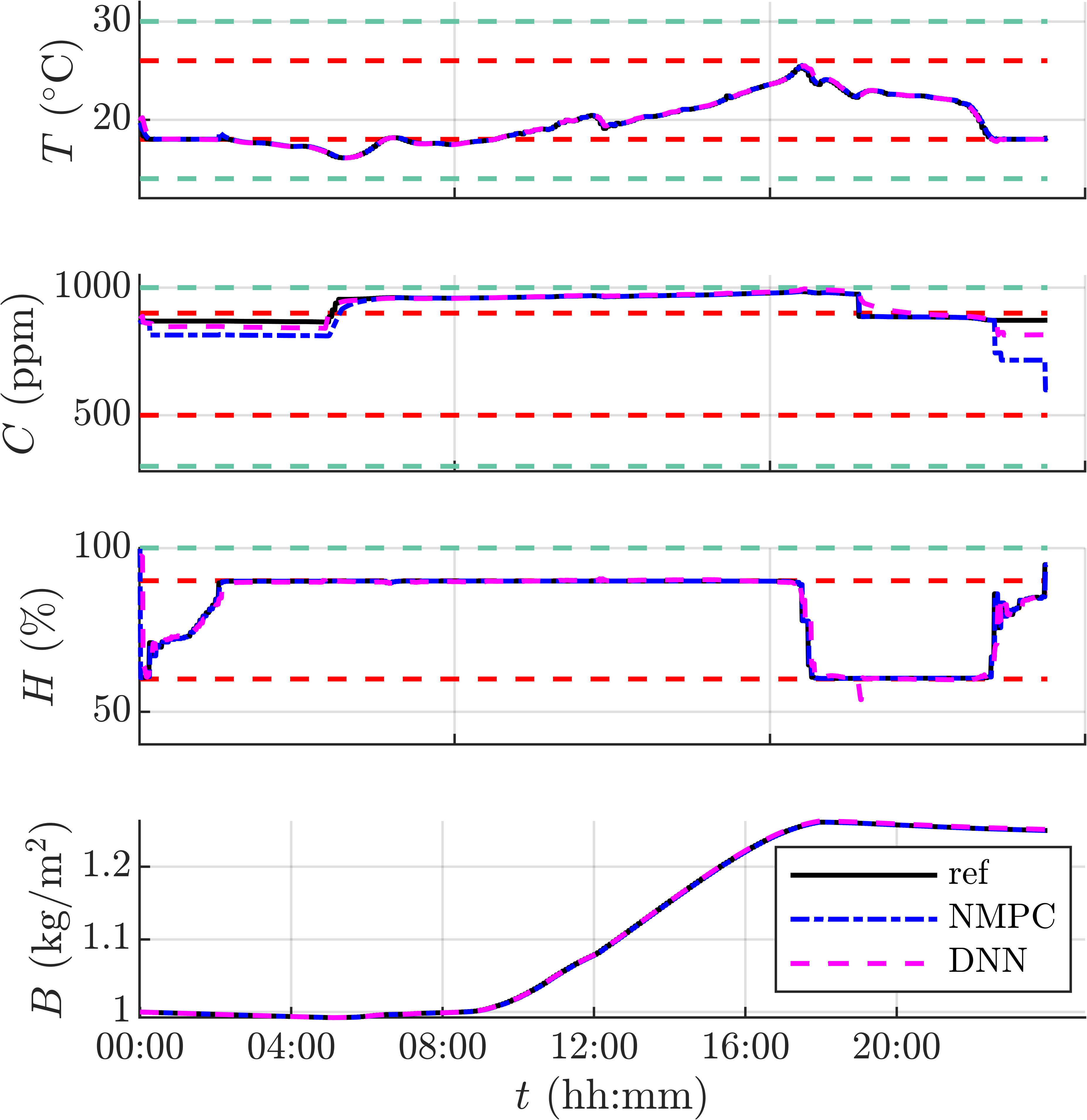} 
	\caption{Tracking performance of NMPC vs. DNN. The red and green dashed lines indicate the soft and hard constraints, respectively. The hard constraint lower bound of $H$ = $10\%$ is not shown here.}
	\label{fig:nonoise}
\end{figure}

\ks{The results of the simulation comparing the tracking performance of NMPC and DNN with respect to the reference trajectory generated by the upper-level OCP for one day are shown in Figure~\ref{fig:nonoise}.}
The initial state vector considered consist of the last real-time state values of the previous day. 
The NMPC optimization as well as DNN inference are performed on a Personal Computer (PC).

\begin{figure}[t!]
	\centering
	\includegraphics[width=8.4cm]{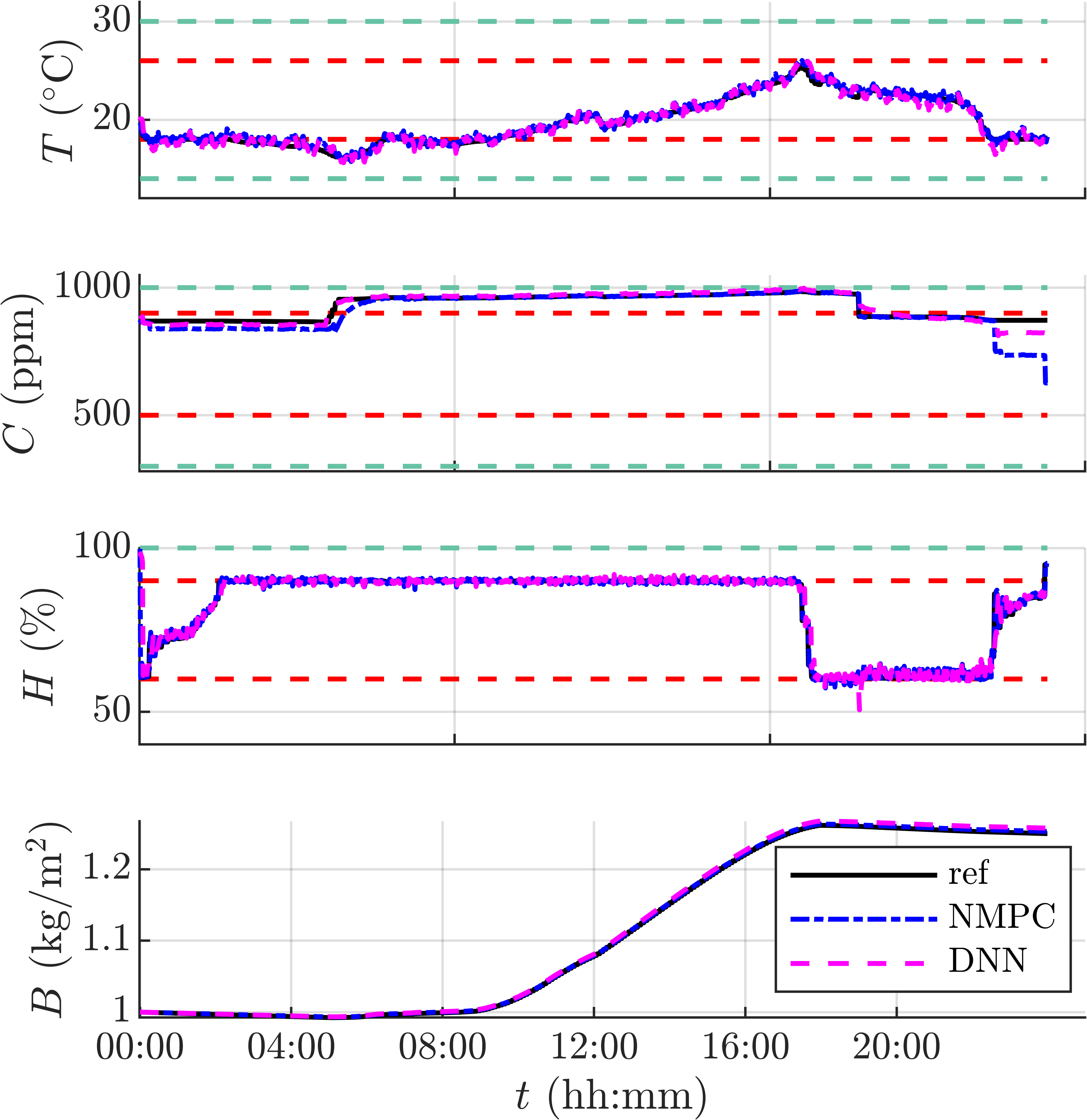} 
	\caption{Comparison of the convergence of states in NMPC vs. DNN under 10$\%$ uncertainty of $\st{d}$ with respect to forecast data. The red and green dashed lines indicate the soft and hard constraints of the states respectively. The hard constraint lower bound of RH - $10\%$ is not shown here.}
	\label{fig:noise}
\end{figure}

First, notice that in both scenarios, the states cross the soft constraints but obey the hard constraints defined by the set~$\mathcal{X}$. 
The CO$_2$ concentration is increased during the onset of solar radiation and maintained at this level. 
The biomass production starts once the temperature reaches around $19 \si{\degreeCelsius}$. 
It can be observed that both the NMPC and DNN are able to track the reference trajectories.
Figure~\ref{fig:noise} shows a comparison of the NMPC and DNN controllers under 10$\%$ uncertainty of $\st{d}$ with respect to the forecast data. 
It can be seen that the performance of the DNN is almost identical to the NMPC controller. 
However, the biomass production by the approximate DNN controller - $1.2497 \, \si{\kilogram \per \meter \squared}$ is slightly higher than the NMPC - $1.2512 \, \si{\kilogram \per \meter \squared}$. 

\ks{The performance of the controllers are examined using the closed-loop net cost $S = V(\st{x}(N_\mathrm{s})) - \sum_i L(\st{x}(i),\st{u}(i))$
	of the greenhouse operation on a day for varying uncertainty in $\st{d}$.
	Table~\ref{tab:compcost} comprises the total costs, operating costs $L$, and terminal cost/income from the crop yield $V$, of the greenhouse operation on a day.}	
\begin{table}[t!]
	\caption{Comparison of net cost of greenhouse operation using NMPC vs. DNN under uncertain disturbance}
	\begin{center}
		\begin{tabular}{ *{7}{c} }
			\toprule
			\multirow{2}{*}{ Uncertainty}  & \multicolumn{3}{c}{ NMPC} & \multicolumn{3}{c}{ DNN} \\
			
			\cmidrule(lr){2-4} \cmidrule(lr){5-7}
			[$\%]$ & S[€] & L[€] & V[€] & S[€] & L[€] & V[€]\\  
			\midrule
			$1\%$ & 113.33 & 12.85 & 126.18 & 100.24 & 12.78 & 113.02 \\ 
			$10\%$ & 68.76 & 30.36 & 99.13 & 103.08 & 13.91 & 117.01 \\ 
			$20\%$ & 76.44 & 32.65 & 109.09 & 101.63 & 16.44 & 118.08 \\ 
			$30\%$ & 70.25 & 38.52 & 108.77 & 99.21 & 19.04 & 118.25 \\ 
			\bottomrule
		\end{tabular}
	\end{center}
	\label{tab:compcost}
\end{table}
\ks{
	\ps{With perfect knowledge of the disturbance~$\st{d}$, total net costs using NMPC are less than the ones with DNN.}
	Alternatively, in the case of uncertainty in $\st{d}$, the DNN outperforms the NMPC.
	The higher operation costs of NMPC indicates the increased control effort to handle $\st{d}$ in the pursuit of reference tracking.  
	Note that the approximate nature of DNN controller and its training with varied disturbance over 8-month training data results in a better performance compared to NMPC.}   

The computation effort of both controllers under different uncertainty levels of $\st{d}$ is shown in Table~\ref{tab:comptime}. 
Compared to NMPC, the average execution time of a DNN controller is faster \ks{by four orders of magnitude for a PC execution.} 

\begin{table}[t!]
	\caption{Comparison of computation time on a PC}
	\begin{center}
		\begin{tabular}{ *{5}{c} }
			\toprule
			\multirow{2}{*}{ Uncertainty[$\%]$}  & \multicolumn{2}{c}{NMPC} 
			& \multicolumn{2}{c}{DNN} \\
			\cmidrule(lr){2-3} \cmidrule(lr){4-5}
			& Mean [\unit{ms}] & SD [\unit{ms}] &  Mean [\unit{ms}] & SD [\unit{ms}] \\  
			\midrule
			$1\%$ & 155.7561 & 74.9314 & 0.0208 & 0.0048 \\ 
			$10\%$& 162.4664 & 87.5212  & 0.0203 & 0.0035\\
			$20\%$ & 159.4011 & 85.8751  & 0.0204 & 0.0028\\
			$30\%$  & 158.1797 & 81.8533  & 0.0227 & 0.0082\\

			\bottomrule
		\end{tabular}
	\end{center}
	\label{tab:comptime}
\end{table}

\section{Conclusion}\label{sec:conclusion}
This paper presents the implementation of a hierarchical model predictive control of a greenhouse with a lower level of fast control using deep neural networks.
Our simulations show a faster computation time and better performance in the presence of uncertain forecast data using a DNN controller compared to NMPC.
The low memory requirement of an approximate DNN controller makes it more desirable for constrained embedded hardware implementation, which results in less power consumption.
\ks{Future work will consider the implementation of the controller using low-cost embedded devices on an experimental setup.}


\bibliographystyle{IEEEtran}
\bibliography{ieee_ecc3}  

\addtolength{\textheight}{-12cm}   









\end{document}